# Experimental observation of nanoscale radiative heat flow due to surface plasmons in graphene and doped silicon

P.J. van Zwol[1*], S. Thiele[1], C. Berger[1,2], W. A. de Heer[2], J. Chevrier[1*]

*1 Institut Néel, CNRS and Universite Joseph Fourier Grenoble, BP 166 38042 Grenoble Cedex 9, France*
*2 School of Physics, Georgia Institute of Technology, Atlanta, GA 30332, USA*

Owing to its two dimensional electronic structure, graphene exhibits many unique properties. One of them is a wave vector and temperature dependent plasmon in the infrared range. Theory predicts that due to these plasmons, graphene can be used as a universal material to enhance nanoscale radiative heat exchange for any dielectric substrate. Here we report on radiative heat transfer experiments between SiC and a $SiO_2$ sphere which have non matching phonon polariton frequencies, and thus only weakly exchange heat in near field. We observed that the heat flux contribution of graphene epitaxially grown on SiC dominates at short distances. The influence of plasmons on radiative heat transfer is further supported with measurements for doped silicon. These results highlight graphene's strong potential in photonic nearfield and energy conversion devices.

*Corresponding authors; petervanzwol@gmail.com,  joel.chevrier@grenoble.cnrs.fr*

As described by Plancks law, farfield (FF) radiative heat transfer (RHT) is a broadband phenomenon. In contrast in near field (NF) [1-21] surface excitations such as phonon polaritons and low frequency plasmons result in an improved spatial coherence [16] and narrow bandwidths [8], leading to an increase in energy density beyond the Planck blackbody limit [1-3]. For materials such as SiC and $SiO_2$ the surface excitations are attributed to ion-vibrations [4,5], whereas for doped silicon [11] and graphene [10, 21] they are due to electronic vibrations (plasmons). For these materials, the plasmon frequency can be tuned by changing the amount of free carriers. In addition for graphene the plasmon frequency $\omega_p$ is tunable by gating, as it depends on the Fermi level [22-24] furthermore the frequency is wave number (q) and temperature dependent [10]. Thus for a given Fermi level one can find a q for which, in the mid IR, there is a plasmon frequency in graphene that matches that of an excitation in another material (fig. 1a). As a result theory predicts that NF RHT is always enhanced when one or both of the dielectric surfaces are covered with graphene [10].

New opportunities in NF have recently emerged for thermo photo voltaics (TPV) [17-20]. It was previously suggested that in NF, surface excitations can enhance the performance of TPV devices beyond the blackbody limit such that larger efficiencies are reached at lower operation temperatures. The almost monochromatic behavior of NF RHT could yield higher energy conversion efficiencies when properly matched to the bandgap of the PV diodes [17-20]. For example, the frequency of plasmons depends on the carrier density, and can therefore be changed by doping, in materials such as silicon. Optimization of TPV efficiency in NF however is difficult due to proximity effects [20]. Thus the RHT enhancing behavior of graphene [21-26] may play an important role there [21]. While previous experiments reported RHT enhancements due to phonon polaritons [4-6], to date no experiments have reported the effect of plasmons or ultra thin films on NF RHT, this is the aim of the present work.

Our experimental setup is a high vacuum interferometric atomic force microscope (AFM) [5]. A bilayer lever that is sensitive to heat flux [4] acts as probe above a plate that is heated with a Peltier element [14]. The schematics of the AFM and the calibration procedure of the bilayer probe have been reported in ref. [14] in NF and FF. The probe used here is the same as characterized in ref [14]. By fitting RHT measurements for a glass sphere and plate with theory, we calibrated the lever sensitivity $S_h=0.037\pm0.008nW/nm$ and the point of contact due to roughness $d_0=66\pm9nm$. In this study we used the following plates: two 205nm silicon on 400nm insulator samples (SoI 205/400, carrier densities $\sim10^{15}$ (SoI15) and $\sim10^{20}cm^{-3}$ (SoI20)), a bare SiC [000$\bar{1}$] sample and two epitaxial graphene on SiC (EG) [000$\bar{1}$] surfaces with 1-2 (EG2) and 5-6 (EG6) graphene layers. Raman spectra at different places on the surface (Witec 632nm) revealed the typical graphene G- and 2D peaks [24] at 1588±7 and 2660±20$cm^{-1}$ respectively, for both EG samples (fig. 1b) (See [24] for Raman spectra where the SiC contribution is subtracted). The thickness of the EG films was estimated with ellipsometry at 15 places on the surface to be $0.5\pm0.05nm$ for EG2 and $1.5\pm0.2nm$ for EG6. Surface roughness was probed with AFM. Our bare SiC sample shows 0.8nm rms roughness with up to 30nm high peaks.

Graphene layers and pleats (of 20nm high) were typically present on EG surfaces (fig. 1d). For our measurements both SiC and the EG samples were put together on the Peltier ensuring equal experimental conditions. The same was done for the two SoI samples.

Our bilayer lever, while it is rotated 90 degrees [14], is also sensitive to an uncontrolled (probably electro-static) force. Note that it is not only unique to our setup [4,14]. Whatever the origin of the force, we found that it can be well characterized and subsequently removed from the RHT signal [14]. For the samples used here, this signal was small when compared to the measured RHT [14] but not negligible. A correction is applied to the measured data at ΔT>0 by subtracting the measured ΔT=0K data. In the supplemental material [28] we show independent measurements and additional analysis of the force measured at ΔT=0 and 40K, that support our claim that it does not change with temperature. We also found no evidence of large temporal or spatial variations of this force. Furthermore we averaged measurements done at multiple spots on the sample. We measured at 7 different locations for both SoI samples, and at 15 different locations for SiC and EG.

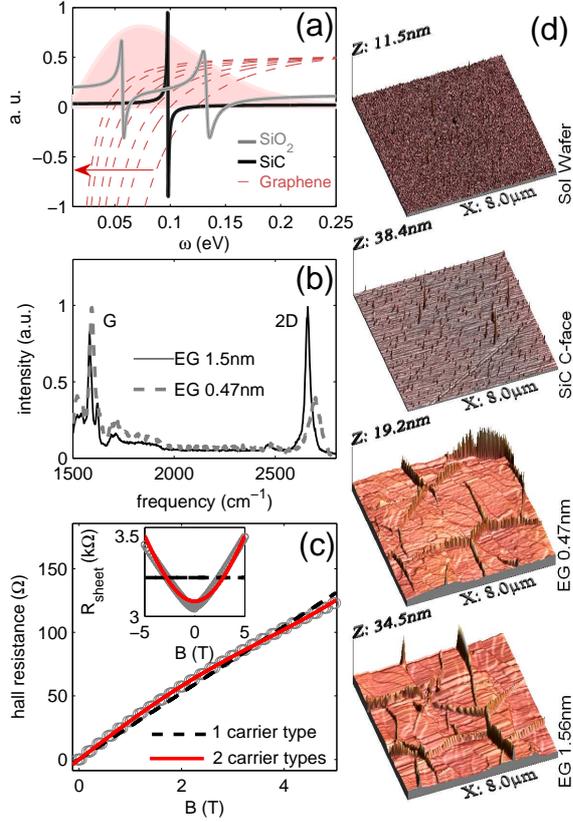

*Figure 1; (a) Dielectric functions of $SiO_2$ and SiC scaled to arbitrary values to compare phonon polariton frequencies to Plancks law at 300K (shaded area) and the q-dependent dielectric function of graphene. The latter is for q in the interval $[2.5-20] \cdot 10^4$ cm$^{-1}$ (arrow), $n_h = 5 \cdot 10^{12}$ cm$^{-2}$ and the relaxation frequency is $10^{13}$ rad/s [10]. (b) Raman spectra for both EG samples. (c) Hall measurements and theory (fits for 1 or 2 carrier types) [25] for the EG2 sample. Sheet resistance is shown in the inset. (d) AFM scans of the samples in this study.*

The efficiency of NF RHT for silicon and graphene to other materials strongly depends on the plasmon frequency [10,11,21], which in turn depends on the amount of free carriers. For the samples used here the carrier density $n_h$ and mobilities $\mu$ were determined by Hall measurements (fig. 1c) in a van der Pauw configuration, with 3mm distance between the contacts, at 4K and 300K using a cryogenic setup and magnetic field up to 5T. For the SoI20 plate the Hall data agreed well for a single type of carriers (holes) with $n_h = 3.6 \cdot 10^{20}$ cm$^{-3}$ and $\mu = 27$ cm$^2$/(V·s). We estimated $n_h = 1.06 \cdot 10^{15}$ cm$^{-3}$ for the SoI15 sample from resistivity measurements, as far as RHT is concerned SoI15 behaves as intrinsic silicon. The non linear Hall resistance as a function of magnetic field measured in graphene, and significant magneto resistance (fig. 1c) suggest that our EG samples have multiple types of carriers (contribution from the several layers). One could use the classical conductivity tensor method like in [25] to estimate an average $n_h$ and $\mu$ of these carriers in each of the graphene layers. However there are multiple solutions to these equations. A simple 2-carrier systems describe the Hall data well, with values in line with the literature for EG2 and EG6 for $n_h \sim 1.5-4 \cdot 10^{12}$ cm$^{-2}$ and $\mu \sim 800$ cm$^2$/(V·s). We stress that the Hall bar of 3mm is fairly large and that the result integrate inhomogeneities, defects, and spatial variation in $n_h$, $\mu$ and the number of EG layers, which complicates this analysis, and makes it unsuitable for use in calculations that are compared to local RHT measurements on microscale areas. It is generally accepted that most carriers in EG are n-type and reside in the layer closest to SiC with $n_h = 2-4 \cdot 10^{12}$ cm$^{-2}$ and $\mu = 1000-10000$ cm$^2$/(V·s) [23-26]. Inner layers have very low carrier densities ($n_h < \cdot 10^{10}$ cm$^{-2}$). The top layer is likely to be almost undoped as well because our setup operates in high vacuum and $T_{sample}$ can reach 375K [27]. Since layers with low carrier density barely contribute to the RHT, we can treat our EG samples as a single conducting layer. Therefore to analyze our RHT data, we will use the present Hall results for doped silicon, and rely on the literature data on micron size Hall bars for EG.

We used standard stochastic electrodynamics [1] to calculate RHT between the two bodies and subsequently used the Derjaguin approximation to obtain results for our plate-sphere system [29]. It is known that this approximation predicts RHT well in NF [14], but underestimates RHT in FF [14,15]. In the case of the SoI15 and SoI20 samples we used multilayer reflection coefficients in the RHT theory. For the SoI20 sample, the carrier scattering time was derived from $\mu$. Together with $n_h$ this defines the Drude term for doped silicon [11]. SiC was optically modeled as a single oscillator [8], and optical data for graphite, silica and sodalime glasses were used from refs. [30-33]. The RHT theory for graphene is based on [10], in which values for $n_h$ and $\mu$ from the literature are used.

In our plane-sphere case the FF RHT was much larger than the NF RHT increase with distance (compare fig. 2a and 2d). For this reason experiment was compared to theory separately for the NF and FF regimes. For all the measurements below, we used the calibration values, $d_0$ and $S_h$ as obtained by fitting RHT theory to experiment for the glass-glass measurements reported earlier [14]. The NF RHT-distance curves were shifted vertically so that at 3μm the values are zero for both theory and experiment, and the FF contribution is effectively subtracted from the NF curves. Then the measurements were shifted horizontally to $d_0$ and theory was scaled with $S_h$.

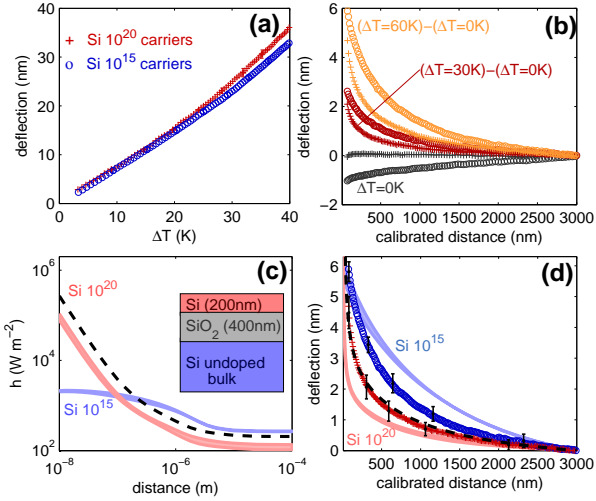
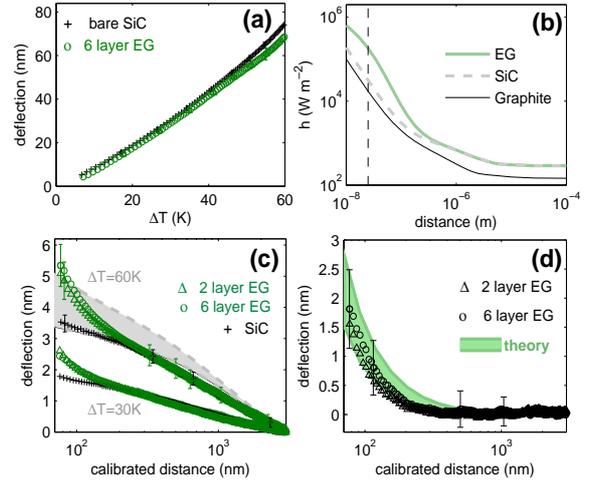

*Figure 2: (a) Measurements of FF RHT for both SoI samples. (b) NF RHT measurements for both SoI samples (see legend symbols 2a). Lever deflection is shown for each ΔT. For curves with ΔT>0 the ΔT=0 curve is subtracted. (c) Plane-plane RHT theory in which the Hall results are used, between $SiO_2$ and both SoI samples. For the latter the multilayer structure is shown. The dashed line represents the optimal Drude parameters for which NF RHT theory matches best the experimental results. (d) Measured average and standard deviation due to spatial variation of NF RHT at 60K as compared to theory (legend in fig. 2c for theory and fig. 2a for experiment applies).*

*Figure 3: (a) Measured FF RHT. (b) Plane-plane theory between SiO2 and SiC, bulk graphite or EG (single layer $n_h=2·10^{12}cm^{-2}$, $\mu=5000\ cm^2/(V·s)$) with plate temperatures 300 and 360K. Theory [10] is valid above 25nm (vertical dashed line). (c) Average and standard deviation due to spatial variations, of NF RHT measurements. Theory (grey area) for SiC-silica is also shown in the plane-sphere case. (d) Experimental and theoretical difference in NF RHT between EG-glass and SiC-Glass for ΔT=60K. Here the optical properties of glass matter less in the theory.*

Figure 2 depicts measurements and theory for RHT measured between the $SiO_2$ sphere and the SoI plates. In FF we found that RHT was almost the same for the two SoI samples (fig. 2a). The measured FF RHT was 25% lower for $SiO_2$-Silicon than for $SiO_2$-$SiO_2$ (see ref. 14), in good agreement with theory. In NF our measurements (fig. 2b) indicate a reduced RHT from plate to sphere for the doped SoI20 sample as compared to the SoI15 sample. We found that RHT increased more rapidly with decreased distance for the SoI20 sample, which is indeed due to the surface plasmon as predicted by theory (fig. 2c). In the plane sphere case RHT is integrated over the sphere area and therefore varying distances. For the present sphere the distance regime around 1μm has a dominant effect. At 1μm, between parallel plates, RHT is larger for the SoI15 sample (fig. 2c). Nonetheless theory and experiment are in reasonable agreement and for both samples and the general behavior is well reproduced (fig. 2d).

Uncertainty in the theory is depicted by plotting the RHT results as an area instead of a line. For the SoI15 sample we used different dielectric data for silica [31-33]. For the SoI20 sample we varied $n_h$ by a factor two around the measured value from *1.8 to 7.2 $10^{20}cm^{-3}$* with corresponding scattering times of *15.5-3.7fsec* as obtained from $\mu$ [11]. The optimal parameters for the SoI20 sample were $n_h=1.3·10^{20}cm^{-3}$ with a relaxation time of *10fsec* (fig. 2c,d), which is close to the Hall results. With these values, there is an excellent agreement between theory and measurement in NF (dashed line fig. 2d), and calculated FF underestimates experiment by 20% (figs. 2a,c). Apart from experimental errors in the Hall and RHT measurements additional uncertainty in the calculated curve may come from the use of tabulated optical properties for the 400nm buried oxide (which was modeled as bulk silica). Note that FF RHT is also much more affected by the oxide multilayer structure. Calculations indicated that the effect of the multilayer structure (this thin film effect as compared to bulk Si is not shown in fig. 2c) yielded a 15% increase in FF RHT for the SoI20 sample and a 15% decrease for the SoI15 sample as compared to bulk silicon, while the RHT increase in NF was hardly affected. For doped silicon we could obtain very good agreement with theory by slightly varying $n_h$ around the measured values. But for intrinsic silicon there were no parameters in the theory that could be varied. We suspect that the Derjaguin approximation may be less accurate here because the NF RHT is barely distant dependent in this case (fig. 2c, [28]).

Results for EG are shown in figure 3. In FF we found that graphene had almost no effect on the RHT (fig.

3a). In NF it was clear from each individual measurement curve that, graphene enhanced the RHT significantly below 200nm (fig. 3c,d). This means that the RHT has increased considerably for the area of the sphere closest to the graphene surface (fig. 3b). Note that, on the contrary, RHT would have decreased for bulk graphite (fig. 3b). Furthermore we found no measurable difference in NF RHT between the EG2 and EG6 samples as seen in fig. 3c,d. This result is in agreement with a single layer analysis, and the assumption that the doped graphene layer at the interface dominates.

The calculated NF RHT for $SiO_2$-SiC in a plate-plate configuration was found to vary by a factor of 5 for different sets of dielectric data for $SiO_2$ [28, 31-33]. In the plate-sphere case for $SiO_2$-SiC the calculations varied significantly but only within 20%. For calculations based on the dielectric data of sodalime glass [32,33] the results were 20% higher than the experimental result. The precise result strongly depended on the absorption of silica in the range of the phonon-polariton peak in SiC. Additional uncertainty may come from the use of an oscillator model for SiC.

By taking the difference between the experimental curves with and without graphene, we extracted the effect of graphene on the NF RHT. In fig. 3d, NF RHT calculations for graphene, based on ref. [10] with values for $n_h=[2,4]\cdot 10^{12}$ $cm^{-2}$ and $\mu=[1000,10000]$ $cm^2/(V\cdot s)$ as found from the literature, are compared to experiment without any other adjustable parameters. In general we found that the RHT calculations were much more sensitive to $n_h$ than on $\mu$. Also $n_h$ and $\mu$ in graphene depend on each other as graphene with lower $n_h$ has higher $\mu$. The effects on RHT calculations offset each other to some extent, and the agreement with RHT experiment remains good for a wide range of $n_h$ and $\mu$ values.

Concluding we have experimentally shown that thermally excited plasmons in doped silicon and EG on SiC substantially increase NF RHT. It is intriguing to see that one or a few atomic layers can completely alter NF RHT and that graphene can enhance RHT for materials with non-matching surface excitations. SiC is often used as a cheap emitter in FF TPV devices, and is one of the most promising candidates for large scale graphene fabrication. EG grown on SiC may therefore serve as a base for future plasmonic NF TPV devices with increased efficiency beyond the blackbody limit.

**Acknowledgements :** We gratefully acknowledge support from the Agence Nationale de la Recherche through the Source-TPV project ANR 2010 BLAN 0928 01, from the National Science Foundation (DMR-0820382), the Air Force Office of Scientific Research and the Partner University Fund. The authors benefited from exchange of ideas within the ESF Research Network CASIMIR. We thank N. Bendiab, L. H. Diez, L. Ranno, F. Gay and J. Marcus for help on Raman and Hall measurements.

*References*

[1] D. Polder and M. Van Hove, Phys. Rev. B 4, 3303 (1971).
[2] C. M. Hargreaves, Phys. Lett. 30 A, 491 (1969).
[3] A. Kittel, W. Müller-Hirsch, J. Parisi, S. A. Biehs, D. Reddig, M. Holthaus, Phys.Rev. Lett. 95, 224301 (2005).
[4] S. Shen, A. Narayanaswamy, and G. Chen, Nano Lett. 9, 2909 (2009). S.Shen, A. Mavrokefalos, P.Sambegoro, G. Chen, Appl. Phys. Lett. 100, 233114 (2012).
[5] E. Rousseau, A. Siria, G. Jourdan, S. Volz, F. Comin, J. Chevrier, and J-J. Greffet, Nat. Photonics 3, 514 (2009).
[6] R. S. Ottens, V. Quetschke, Stacy Wise, A. A. Alemi, R. Lundock, G. Mueller, D. H. Reitze, D. B. Tanner, and B. F. Whiting, Phys. Rev. Lett. 107, 014301 (2011).
[7] T. Kralik, P. Hanzelka, V. Musilova, A. Srnka, and M. Zobac, Rev. Sci. Instrum. 82, 055106 (2011).
[8] J.-P. Mulet, K. Joulain, R. Carminati, and J.-J. Greffet, Appl. Phys. Lett. 78, 2931 (2001).
[9] I. Volokitin and B. N. J. Persson, Rev. Mod. Phys. 79, 1291 (2007), ibid. Phys. Rev. B 83, 241407(R) (2011).
[10] V. B. Svetovoy, P.J. van Zwol, J. Chevrier, Phys. Rev. B 85, 155418 (2012)
[11] C. J. Fu and Z. M. Zhang, Int. J. Heat Mass Transfer 49, 1703 (2006).
[12] M. Francoeur, M. P. Mengüç, and R. Vaillon, Appl. Phys. Lett. 93, 043109 (2008).
[13] S.-A. Biehs and J.-J. Greffet Phys. Rev. B 82, 245410 (2010).
[14] P. J. van Zwol, L. Ranno, J. Chevrier, J. Appl. Phys, 111, 063110 (2012), Ibid, Phys. Rev. Lett. 108, 234301 (2012).
[15] C. Otey, S. Fan, Phys. Rev. B 84, 245431 (2011).
[16] J. -J. Greffet, R. Carminati, K. Joulain, J. -P. Mulet, S. Mainguy, and Y. Chenet, Nature (London) 416, 61 (2002).
[17] R. S. DiMatteo, P. Greiff, S. L. Finberg, K. A. Young-Whaite, H. K. H. Choy, M. M. Masaki, and C. G. Fonstad, Appl. Phys. Lett. 79, 1894 (2001).
[18] M. D. Whale and E. G. Cravalho, IEEE Trans. Energy Convers. 17, 130 (2002).
[19] A. Narayanaswamy and G. Chen, Appl. Phys. Lett. 82, 3544 (2003).
[20] M. Laroche, R. Carminati, and J.-J. Greffet, J. Appl. Phys. 100, 063704 (2006).
[21] O. Ilic, M. Jablan, J. D. Joannopoulos, I. Celanovic, and M. Soljačić, Optics Express, 20, A366-A384 (2012).
[22] K. S. Novoselov, A. K. Geim, S. V. Morozov, D. Jiang, Y. Zhang, S. V. Dubonos, I. V. Gregorieva, and A. A. Firsov, Science 306, 666 (2004).
[23] C. Berger et al., Science 312, 1191 (2006).
[24] W. A. de Heer, C. Berger, M. Ruan, M. Sprinkle, X. Li, Y. Hu, B. Zhang, J. Hankinson, and E. Conrad, PNAS 108, 16900 (2011).
[25] Y. Lin et al, Appl. Phys. Lett. 97, 112107 (2010).
[26] J. L. Tedesco, B. L. VanMil, R. L. Myers-Ward, J. M. McCrate, S. A. Kitt, P. M. Campbell, G. G. Jernigan, J. C. Culbertson, J. C. R. Eddy, and D. K. Gaskill, Appl. Phys. Lett. 95, 122102 (2009).
[27] M. Sprinkle, D. Siegel, Y. Hu, J. Hicks, P. Soukiassian, A. Tejeda, A. Taleb-Ibrahimi, P. Le Fèvre, F. Bertran, S. Vizzini, H. Enriquez, S. Chiang, C. Berger, W.A. de Heer, A. Lanzara, E.H. Conrad, Phys. Rev. Lett., 103, 226803 (2009).
[28] See supplemental material…
[29] B. V. Derjaguin, Kolloid-Z. 69, 155 (1934).
[30] H.R. Philipp, Phys. Rev. B 16, 2896 (1977).
[31] R. Kitamura, L. Pilon, and M. Jonasz, Appl. Opt. 46, 8118 (2007).
[32] P. J. van Zwol, G. Palasantzas, and J. Th. M. DeHosson, Phys. Rev. E 79, 041605 (2009).
[33] B.G. Bagley, et al. Crys. Solids. 22 423, (1976)

# Supplemental to 'Experimental observation of nanoscale radiative heat flow due to surface plasmons in graphene and doped silicon'

P.J. van Zwol[1], S. Thiele[1], C. Berger[1,2], W. A. de Heer[2], J. Chevrier[1]

*1 Institut Néel, CNRS and Universite Joseph Fourier Grenoble, BP 166 38042 Grenoble Cedex 9, France*
*2 School of Physics, Georgia Institute of Technology, Atlanta, GA 30332, USA.*

**Electrostatic force signal compared to the radiative heat transfer signal**

When a measurement is done at ΔT=0K, we observe a lever deflection that varies with sphere-plane distance. In our setup [1,2] this deflection is most often small and always of opposite sign to the deflection due to heat transfer. The signal is most likely of electrostatic origin and we may refer to it in that way. It exact origins do not matter to us because we show in addition that;

- A measurement of the force in a measurement configuration where RHT is negligible reveals no changes with temperature in the range 300-340K.

- A change in strength of the force with ΔT cannot explain the observed effects attributed to RHT and plasmons

Note at last that the ΔT=0 signals [1] were essentially constant from place to place as measured over the timeframe of a week (fig. 1). Note also that in the RHT setup the lever is perpendicular to the surface [2] to reduce the effect of forces, but even in that case a net force may be observed due to a torque. We have done an additional test of the force with the lever parallel to the surface (fig. 2). As such there is maximum force sensitivity for the lever and the force overwhelms the heat transfer signal by about two orders of magnitude. This measurement was done with a lever of the same type, and same sized glass sphere (Veeco MLCT 320μm, and 40μm sphere) as the one used for the RHT measurements. The observed deflections due to force are about 10-500 times larger than the electrostatic forces observed for all materials investigated with our RHT setup with the lever perpendicular to the surface [2]. From simple geometrical arguments [2] one can estimate that a factor 10-100 is to be expected when changing the lever from perpendicular to parallel with respect to the surface if one considers the torque to be a sine component of the force due to a positioning error of 1-3 degrees [2]. In addition there may be somewhat more charge on the particular sphere used for fig. 2 below. We saw no changes in the deflection signal due to forces in the temperature range 300,340K for this sphere. This is in agreement with what we have claimed before.

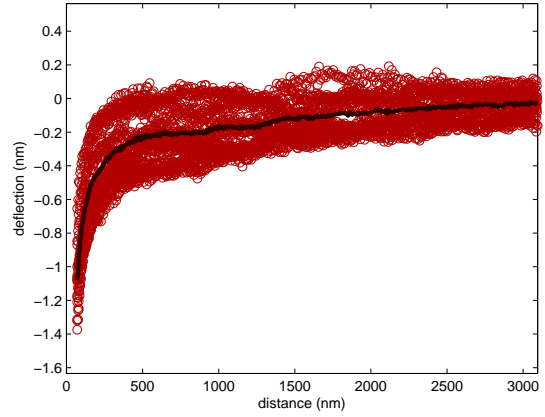

*Figure 1: Electrostatic force or ΔT=0 signal for SiC measured at 15 different places. This data is obtained over the course of a week. Apart from noise no large variations are seen in this timeframe. The electrostatic force appears constant in time and from place to place.*

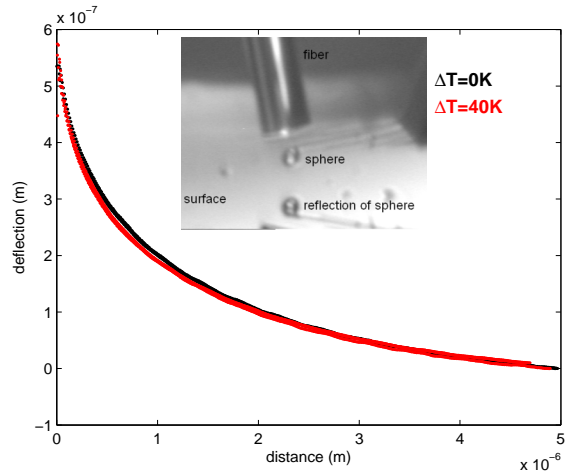

*Figure 2: Electrostatic force measured in high vacuum for the configuration shown in the inset between glass and the SoI15 sample (compare to inset in fig.4 in this supplemental). The deflection due to force is about 100 times larger than deflection due to that of radiative heat transfer in this case (compare to fig. 3a). The force constant of the lever is 0.01N/m. No significant change in force is observed within the specified temperature range. We obtained the same result repeatedly on different locations on the surface.*

Note that lightly doped silicon may decrease its resistance by a factor of 30 from T=300K to T=360K [4], yet we have no evidence that the electrostatic force between lightly doped silicon changes in this range. For insulating materials such as glass no measurable change in resistance occurs below the glass temperature [3]. We measured that the resistance of graphene decreased by 3% in this temperature range. Finally the resistance for heavily doped silicon decreases by 15% in this temperature range [4]. Only for $VO_2$ we have evidence that the electrostatic force changes at the metal insulator transition, but for this sample the resistance changes with 4 orders of magnitude [2].

We can ask further how the electrostatic force must change with temperature to explain the observed effects that are already well explained with standard heat transfer theory. Note first, that the observed deflection due to forces is of opposite sign as compared to the deflection due to RHT for all samples we investigated. Thus the forces need to switch sign if they are to explain the observed deflection due to RHT. We then verified whether an eventual increase in electrostatic force can explain the measured RHT differences between SiC and EG and between SoI15 and SoI20. In fig. 3 the electrostatic force is already subtracted from the $\Delta T=30,60$ curves. It is clear from fig. 3b that, for the electrostatic force difference (between SoI15 and SoI20) to explain the observed differences at $\Delta T>0$, it should change its distance scaling from monotonical to nonmonotoinical besides changing its strength by a factor of 2 for a temperature change of 20% (from 300 to 360K). Note that this means that the electrostatics measured for one sample must change its value respective to the other sample by 200% upon heating. Nonmonotonic electrostatic forces between a sphere and a plate do to our knowledge not exist in vacuum or air for the plate-sphere configuration. RHT theory describes the observed nonmonotonic effects with distance very well however. The differences between RHT theory and experiment were discussed in the main text of the manuscript. For EG and SiC (fig. 4) there is a very similar situation. In order for the electrostatic force difference (between SiC and EG) to explain the differences that we attribute to RHT, it would need to change scaling with distance considerably besides changing its strength with $\Delta T$ (a factor 2) for one sample with respect the other. Radiative heat transfer theory provides a much simpler explanation.

.

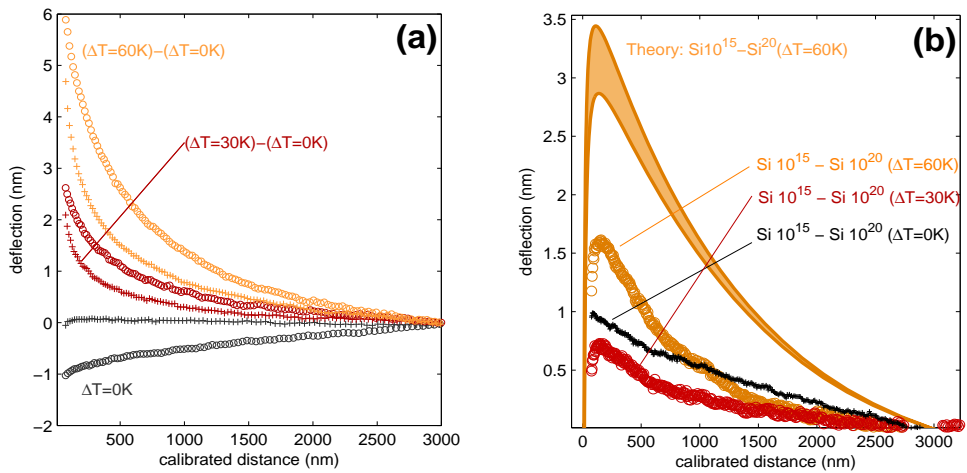

*Figure 3: (a) Plot of fig. 2a of the manuscript. (b) Plot of the differences between the SoI15 and SoI20 samples at all temperatures. For $\Delta T=30,60$ the $\Delta T=0$ curves are subtracted, thus this represents deflection only due to RHT.*

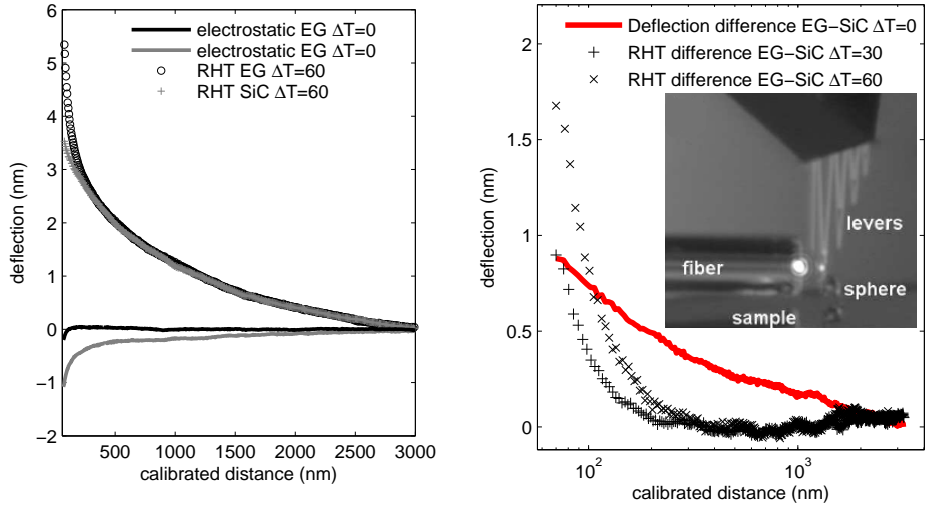

*Figure 4: (a) Averages of all measurement for EG and SiC at ΔT=0 and ΔT=60 (electrostatic ΔT=0 signal is subtracted). (b) Difference in measured deflection between EG and SiC versus distance for ΔT=0,30,60. Again for ΔT=30,60 the ΔT=0 signals are subtracted. The inset shows the configuration [2] of our setup that is used to minimize forces (compare this to that o the inset in f fig. 2 in this supplemental).*

**Dielectric functions of different glasses near the SiC absorption peak**

In figure 5 we show dielectric data for several glasses. The behavior of the different dielectric data for glass lead to a large variation in calculated heat transfer around the SiC phonon polariton peak. The dielectric data is plotted on a log scale, and the variations therein are considerable.

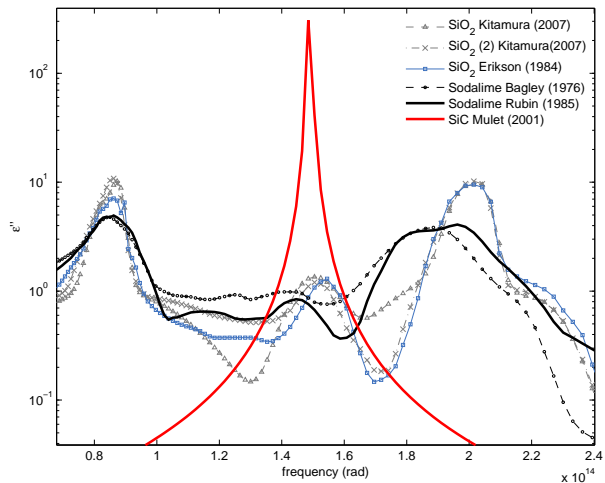

*Figure 5: Imaginary part of the dielectric function for several glasses (ref. [5-9]) around the SiC phonon polariton peak. The two SiO2 phonon polariton peaks are visible too. There is considerable variation in aborption in this range, which affects the radiative heat transfer calculations.*

**Radiative heat transfer for intrinsic silicon**

Silicon is the only material for which we found that experiment deviated from near field RHT theory based on the Derjaguin approximation. For other materials we investigated it worked fine in near field [10]. Our suspicion that the Derjaguin approximation may be less accurate for intrinsic silicon is based on figure 6 below. Note that for all measured materials that the RHT increases strongly with decreasing distance below 1 micron, except silicon. The bulk of the RHT increase with distance for silicon happens in the regime around 1-3 micron. For silicon this range always dominates in the integration over the sphere area even for the plate sphere distances down to 70nm. This isn't the case for all other materials we investigated. This means that the effective interaction length for silicon is about 3 microns no matter how small the distance between the sphere and the plate. This is not negligibly small as compared to the sphere radius.

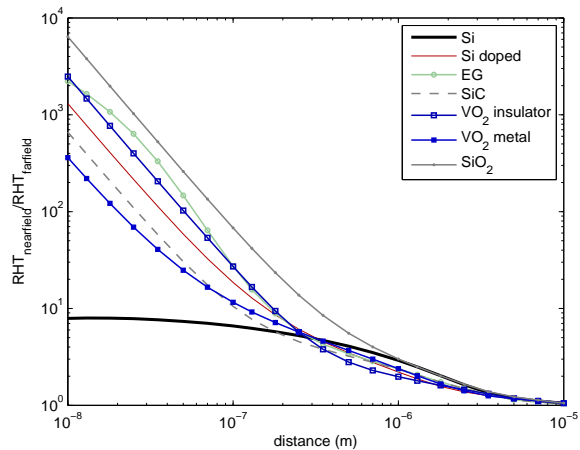

*Figure 6: Normalized radiative heat transfer increase from farfield (=1) in the plate-plate configuration for all materials tested in this work and from ref 2. For all materials we find good correspondence with theory based on the Derjaguin approximation, except for silicon for which we see only very little increase of RHT with distance below 1 micron.*

**References**

[1] S.Shen, A. Mavrokefalos, P.Sambegoro, G. Chen, Appl. Phys. Lett. 100, 233114 (2012).
[2] P. J. van Zwol, L. Ranno, J. Chevrier, J. Appl. Phys, 111, 063110 (2012), Ibid, Phys. Rev. Lett. 108, 234301 (2012).
[3] A.A. Somerville, Phys. Rev. (Series 1) 31 261 (1910)
[4] J. Y. W. Seto, J. App. Phys. 46, 5247 (1975)
[5] R. Kitamura, L. Pilon, and M. Jonasz, Appl. Opt. 46, 8118 (2007).
[6] B.G. Bagley, et al. Crys. Solids. 22 423, (1976)
[7] J.-P. Mulet, K. Joulain, R. Carminati, and J.-J. Greffet, Appl. Phys. Lett. 78, 2931 (2001).
[8] M. Rubin, Sol. Energy Mater. Sol. Cells 12, 275 (1985)
[9] T. S. Eriksson, S. Jiang, and C. G. Granqvist, Applied Optics 24 745 (1985)
[10] C. Otey, S. Fan, Phys. Rev. B 84, 245431 (2011).